Two-photon comb with wavelength conversion and 20-km distribution for quantum communication

Authors

Kazuya Niizeki[1]

Daisuke Yoshida[1]

Ko Ito[1]

Ippei Nakamura[1,2]

Nobuyuki Takei[3]

Kotaro Okamura[4]

Ming-Yang Zheng[5]

Xiu-Ping Xie[5]

Tomoyuki Horikiri[*,1,6]

Affiliations

Yokohama National University[1], 79-5 Tokiwadai, Hodogaya, Yokohama, Kanagawa 240-8501, Japan

KISTEC[2], 705-1 Shimoimaizumi, Ebina, Kanagawa 243-0435, Japan

Kyoto University[3], Kitashirakawa, Oiwake, Sakyo, Kyoto 606-8502, Japan

Kanagawa University[4], 3-27-1 Rokkakubashi, Kanagawa, Yokohama, Kanagawa 221-8686, Japan

Jinan Institute of Quantum Technology[5], Jinan, Shandong 250101, China

JST PRESTO[6], 4-1-8 Honcho, Kawaguchi, Saitama 332-0012, Japan

Abstract

Quantum computing and quantum communication, have been greatly developed in recent years and expected to contribute to quantum internet technologies, including cloud quantum computing and unconditionally secure communication. However, long-distance quantum communication is challenging mainly because of optical fiber losses; quantum repeaters are indispensable for fiber-based transmission because unknown quantum states cannot be amplified with certainty. In this study, we demonstrate a versatile entanglement source in the telecom band for fiber-based quantum internet, which has a narrow linewidth of sub-MHz range, entanglement fidelity of more than 95%, and Bell-state generation even with frequency multimode. Furthermore, after a total distribution length of 20-km in fiber, two-photon correlation is observed with an easily identifiable normalized correlation coefficient, despite the limited bandwidth of the wavelength converter. The presented implementation promises an efficient method for entanglement distribution that is compatible with quantum memory and frequency-multiplexed long-distance quantum communication applications.

Introduction

Recently, quantum technologies have been intensively developed globally. For instance, a quantum processor with 53 qubits was demonstrated by Google [1], and they reported that quantum supremacy was verified experimentally. If even larger numbers of qubits become controllable in the future, Rivest–Shamir–Adleman (RSA) cryptosystems could be decrypted using quantum computers [2]. Quantum communication needs to be developed for the purposes of unconditional security [3][4] and of efficient quantum computing by connecting quantum computers through quantum communication network, or quantum internet [5]. Quantum internet connects remote nodes using quantum entanglement and can enable technologies such as quantum networks of clocks [6], secure cloud quantum computing [7], and distributed quantum computing. The key element of this technology is quantum entanglement, which is the basis of quantum internet and can be directly applied to quantum teleportation [8], entanglement swapping [9], and dense coding [10].

However, long-distance quantum communication involves challenges due to losses in optical fibers, where unknown quantum states cannot be amplified with certainty, resulting in the need for a quantum repeater for fiber-based transmission. Satellite-based experiments (without a fiber network) have demonstrated 1200-km entanglement generation [11], 7600-km BB84 with a decoy state containing a trusted relay [12], and photon transmission over ~20000 km from a medium-Earth orbit satellite [13]. Although atmospheric transmission can, in principle, facilitate long-distance communication relatively easily, it is significantly limited by air conditions and beam diffraction [12]. In this case, optical-fiber-based transmission can realize more stable operation not only because fibers are not highly dependent on weather conditions but also because a colossal network of fibers has already been installed around the world.

The solution to overcoming the distance limitations is to use quantum repeaters, which mainly consist of Bell-state operators with [14] or without [15] quantum memory (QM). The maximum distance of quantum key distribution (QKD) without using quantum repeaters was estimated to be approximately 550 km [16]; the maximum distance achieved experimentally to date is 421 km [17]. Aiming beyond secure communication of classical information by QKD, for instance, to combine quantum nodes such as quantum simulators and/or quantum computers, QM is required [18] to assist the repeater and preserve quantum information with arbitrary operation timing for multi-partite processing.

Based on the above discussion, QM-compatible versatile entanglement source (VES) is ardently desired to approach long-distance fiber-based quantum internet. The VES should emit telecom wavelength photons (∼1.5 μm) to minimize fiber loss, have a sufficiently narrow linewidth for many types of QM ($\ll 10\,\mathrm{MHz}$), and achieve high-fidelity quantum entanglement. Each of these specifications, however, is very difficult to realize that there have been only a few reports [19] about VES to date: It is still challenging to obtain all of high photon-count rate, narrow linewidth and high entanglement fidelity in the telecom regime. In a previous study, cavity-enhanced spontaneous parametric down conversion (SPDC) photon source was demonstrated for many applications [20]; for example, demonstrating theoretical/proof-of-principle models [21]–[26], highly bright single modes [27]–[31], and narrow linewidths [32]–[35]. On the other hand, for compatibility between telecom and QM absorption line, wavelength conversion (WC) is demonstrated using a laser [36][37] and single photons [38][39], for application to nitrogen-vacancy centers, rare-earth-doped crystals, and rubidium gas, among others. WC still has difficulty about conversion efficiency and noise photon to realize photonic interface.

In this study, we demonstrate a VES with wavelength conversion after 10-km fiber transmission. We utilize the two-photon comb (TPC) technique, which realizes a large number of frequency multimodes and entangled photon pairs with a narrow sub-MHz linewidth even in a telecom band. The WC is based on sum-frequency generation (SFG) and target $Pr^{3+}$:YSO QM. We successfully realize two-photon 10-km transmission in an optical fiber or overall 20-km distribution and subsequent WC, resulting in a clear observation of comb structure of wavelength-converted TPC (WC-TPC), which is most suitable for use in frequency multiplexed quantum communication. This technical development can be applied not only to quantum information science but also to experimental optics.

Results

*TPC specifications*

The TPC used herein consists of degenerate 1514-nm SPDC crystals and a surrounding cavity. Further, SPDC occurred in type-0 periodically poled lithium niobite (PPLN) with temperature stability on the order of a few millikelvins. As shown in **Fig. 1**, we used a mutually orthogonal arrangement to generate two-photon polarization entanglement $\alpha|HH\rangle + \beta|VV\rangle$, where $\alpha$ and $\beta$ are the probability amplitudes including relative phase [40]. This type of entanglement generation method has three advantages: high brightness or photon-generation rate compared with other phase matching types based on overlapped SPDC cones [40], low dependence on optics alignment, and path/phase compensation caused by birefringence in the crystal. However, this type of entanglement generation reduces the finesse of the surrounding cavity because of optical losses (e.g., via absorption, scattering, and reflection) in the crystal, which is a serious problem related to the absorption line of QM. For example, when two 1-cm-long PPLN crystals are placed inside a typical 0.5-m cavity, the theoretical linewidth of the cavity is ~5.3 MHz with a typical PPLN loss of 0.06 dB/cm [41] in the impedance-matched case. Photons with such linewidth will be coupled to a 10-MHz-linewidth QM with an efficiency of ~70% (we assume Lorentzian photon profile and squarish QM absorption line). Furthermore, because this value can only be obtained in a perfectly aligned case, careful cavity alignment is always required, which requires an additional complicated maintenance procedure. To overcome this disadvantage, we developed a ~2.5-m-long bow-tie cavity with a free spectral range (FSR) of ~120 MHz, which enabled a sub-MHz linewidth along with an increase in the number of frequency modes. All frequency modes are entangled in each photon pair and have good compatibility with atomic-frequency-comb QM, which enables time- and frequency-multiplexed quantum communication [42]. Cavity locking was achieved using the Pound–Drever–Hall technique to target the absorption line of QM and stabilize the relative phase of entanglement.

To evaluate the quality of the generated two photons, we measured two-photon statistics with a Hanbury Brown–Twiss-type setup, which consists of a 50:50 beam splitter, two superconducting single photon detector (SSPDs), and a time-correlated single-photon counting module (TCSPC). We utilized SSPD and silicon avalanche photodiode (SiAPD) depending on whether the wavelength of photon was telecom or visible. The SSPDs have efficiencies of approximately 85% at telecom wavelength; **Fig. 2** (a) shows the raw data of two-photon correlation with 10-μW pump power. This type of Glauber's correlation function, which is similar to a comb structure, indicates frequency multimodality as in the Fourier relation. The measured time interval, corresponding to the time taken for a round trip in the cavity, was 8.6 ns, and the FSR was 116 MHz. In general, the ratio between the time interval and the pulse width of each peak gives the number of frequency modes or comb range: if some fiber dispersion exists, the number of modes can be evaluated because the pulse width becomes substantially larger than the timing jitter and thus the timing jitter becomes ignorable. Then, we performed 10-km fiber transmission and evaluated the width of the frequency range by measuring the dispersion. We observed ~2 ns broadening per one channel of TCSPC, and we estimated that TPC had a comb range of approximately 1–2 THz from the fiber dispersion of 15 ps·nm$^{-1}$·km$^{-1}$ (please see Supplementary note 3).

The exponential envelope with a long coherence time implies a narrow cavity linewidth $\Delta v$ that is affected by the loss over the round trip. By approximating the envelope as $e^{-2\pi(\Delta v)t}$, we calculated $\Delta v$ as 0.95 MHz (or, according to Ref. [25], the degenerate photon linewidth will be 0.64 times this value, i.e., ~0.61 MHz). Whereas this is the case with an output mirror of 99% reflectivity, an output mirror of 95% reflectivity resulted in a broader cavity linewidth of 1.35 MHz and a higher brightness because the escape efficiency was approximately 2 to 3 times higher.

Photonic-state tomography was performed using the mirror of 95% reflectivity to obtain more precise counts in a shorter acquisition time; **Fig. 2** (b) and (c) show the reconstructed density matrix of the absolute value after applying the maximum-likelihood method to 16 measurements [43]. The maximal fidelity to an arbitrary pure state was 96.1% and the concurrence was 93.0%, which were the highest values in the multimode regime (please see Supplementary note 6). Through 16 similar measurements, we further obtained a Clauser–Horne–Shimony–Holt parameter [44] $S$ of 2.47 (> 2 implies nonlocality). Subsequently, we placed zero-order half-wave plates on the path after the beam splitter to adjust the relative phase of the entangled state and form a Bell state. Four Bell states were achieved (**Fig. 3**) by changing the angle of the horizontal plane, i.e., the yaw angle and slow axis in the vertical plane. The fidelity to $|\Phi^+\rangle = |HH\rangle + |VV\rangle$, $|\Phi^-\rangle = |HH\rangle - |VV\rangle$, $|\Psi^+\rangle = |HV\rangle + |VH\rangle$, and $|\Psi^-\rangle = |HV\rangle - |VH\rangle$ was 90.0%, 90.2%, 89.4%, and 88.1%, respectively (we omitted the coefficients $1/\sqrt{2}$). These results show that this technique is effective, even for frequency-multiplexed entanglement.

*Wavelength-converted two-photon comb*

The principle of WC is sum frequency generation (SFG) in a PPLN waveguide with a strong auxiliary laser. Our target wavelength was ~606 nm, which is the center of the Pr$^{3+}$:YSO absorption line $^3H_4(0)$-$^1D_2(0)$ [45]. In the whole experiment, we used 1514-nm photons and a 1010-nm laser, which satisfy the following two conditions: their sum frequency must be 606 nm and they can be stabilized by molecular gases of acetylene and iodine, respectively. Our setup aimed at conversion of a time-bin state which is suitable for long-distance fiber transmission and would achieve proper quantum frequency conversion because the temporal/linewidth profile of input photons would be preserved. **Fig. 4** (a) shows WC-TPC with a SPDC pump power of 10 mW and SFG laser power of 50 mW. The WC-TPC has two important parameters: noise floor level and signal-to-noise ratio (SNR). The noise floor level is influenced by many factors, including the presence of a residual laser, which can induce other nonlinear processes such as SPDC and Raman scattering [36][37], dark counts in the detector, and stray light (related results are presented in the Supplementary note 5).

Two-photon correlation was characterized by the normalized second-order signal-idler correlation coefficient $g_{s,i}^{(2)}(0)$, which is defined as the ratio of the highest count to the average noise count, similar to SNR (**Fig. 4** (b)). The blue circles connected by a line represent $g_{s,i}^{(2)}(0)$ before WC (1514-nm two-photon data without WC), and the orange squares represent the values after WC-TPC (the log scale one is in Supplementary note 2). Some scattering was observed in the values after WC-TPC because the timing jitter increased by one order in the visible range compared with the telecom range. The variation in $g_{s,i}^{(2)}(0)$ clearly showed that the low SNR in the signal from the two photons in the telecom range can be increased by putting the two photons through the wavelength converter (see Discussion for details). When we tried to demonstrate only single-photon WC, which corresponds to a correlation between the telecom photon and the visible-range photon, we did not observe a clear correlation function, despite extensive adjustment of the SPDC/SFG pump power and spectral filters; we observed $g_{s,i}^{(2)}(0) \approx 1$, i.e., almost the noise floor (data not shown).

*Wavelength conversion after fiber transmission*

For achieving memory-assisted quantum communication, TPC and WC were combined in an attempt to achieve long-distance communication, assuming that each wavelength has a distinct advantage: telecom qubits can be transmitted across long distances through a fiber with an attenuation of 0.2 dB/km, whereas the visible-range qubits can interact with highly efficient QM. It is very important to ensure that the whole system is applicable to fiber-based quantum communication because other problems such as the degree of polarization mixing and modulation by wavelength dispersion exist.

**Fig. 4** (c) shows WC-TPC after telecom two-photon 10-km transmission; this result is comparable to **Fig. 4** (a), which has the same experimental conditions, except for the fiber length. Even two photons with a total separation of 20 km and a long timing jitter of the detector can yield a $g_{s,i}^{(2)}(0)$ value of approximately 3. The factors decreasing $g_{s,i}^{(2)}(0)$ do not affect the noise floor (but only weaken the coincidence counts) and include fiber losses depending on its length (including a connection loss), polarization rotation induced by birefringence due to the polarization-sensitive WC, and pulse broadening by wavelength dispersion over the time-bin size of the TCSPC. The first factor can be reduced at telecom wavelengths, and our measured transmittance value was approximately 62% at 1514 nm. The second factor can be corrected only when the environmental conditions remain almost unchanged. The third factor is very small because the wavelength converter has a bandwidth of ~0.03 nm. The fiber dispersion of 15 ps·nm$^{-1}$·km$^{-1}$ therefore induces a pulse spreading of 4.5 ps which is much smaller than the timing jitter. For further increase in $g_{s,i}^{(2)}(0)$ with or without a long fiber, the most effective approach is to develop a filter with a suitable frequency range to remove WC noise from the WC-TPC spectra; this technique is discussed in the following section.

Discussion

The TPC has a wide spectral range of ~1 THz, but a very narrow linewidth of ~1 MHz, and it shows polarization entanglement with a fidelity of ~90% to an arbitrary Bell state. In addition, it has a very long time interval and ensures high-speed modulation or time-bin state generation. A narrow linewidth is guaranteed by the long cavity length (or short FSR), even with low finesse. Therefore, the highest finesse in the current setup is not required because the linewidth will remain narrower than the absorption line of QM, even if the finesse reduces slightly due to either a shift in alignment or something else; in our case, the photon with 1-MHz linewidth has a redundancy and will couple to 4.6-MHz window of $Pr^{3+}$:YSO QM even if finesse becomes less than a half of the present value. Accordingly, TPC is free of frequent cavity alignment, resulting in good compatibility with the troublesome Bell-state generation and connection to QM. At smaller finesse, it has been reported that the intensity of side (or additional) clusters of the main frequency comb increases [28]; however, because of their spectral separation, the side clusters can be removed using a spectral filter or can be used to increase the number of modes without filtering. The cavity condition is affected by the temperature and convection of air, which can be compensated for by ensuring that the temperature stability of the crystal is of the order of several millikelvins over a period of few days. The condition of the crystal is the main factor preventing high fidelity because many frequency and time modes exist. The relative phase and coherence will clearly deteriorate if the crystal position is changed. To approach the best condition, we used a polyethylene-terephthalate board to eliminate thermal interaction between the crystal holder and the positioning stage, and a wind shield to reduce the changes in air conditions and cavity length. For further improvement, an external compensation crystal, similar to that used for single-pass SPDC [46], and separative adjustment of the signal/idler using a conjoint double cavity [47] could be included.

To ensure good compatibility between QM and TPC, the QM requires an atomic frequency comb with a tailored absorption line composed of a finely arranged comb-like structure in the wide inhomogeneous-broadening of the rare-earth-ion ensemble (see Supplementary note 7). This QM has many degrees of freedom of frequency where the absorption line is constructed, except in the case of an extremely small frequency difference (~ 10 MHz) because of the hyperfine structure. Some previous studies have investigated multi-frequency modes [48][49]. In the case of $Pr^{3+}$:YSO, the achievable atomic-frequency-comb range is less than 4.6 MHz, and the inhomogeneous bandwidth is of the order of several to tens of GHz (when the concentration of the dopant $Pr^{3+}$ is 1%); however, we could not measure the largest bandwidth accurately. Such a wide inhomogeneous broadening can show good connectivity with the overall comb range of WC-TPC limited by the bandwidth of the wavelength converter. Further, assuming that the profile of a photon linewidth of ~1 MHz is pure Lorentzian and that one of the absorption range of 4.6 MHz is squarish, the coupling efficiency is greater than 90%.

Based on the tomographic results, we consider that the dispersion or wavelength-dependent retardance of the waveplates affects the tomographic reconstruction. To realize desired states, additional waveplates unrelated to the tomography are utilized: one for $|\Phi^\pm\rangle$, and two for $|\Psi^\pm\rangle$. Although estimating the magnitude of the effect is difficult due to complex system, the obtained fidelity does appear lower when this is taken into account.

The increase in $g^{(2)}_{s,i}(0)$ after the WC was attributed to the different sources of noise before and after the WC, and it also indicates a post-selective effect in which an inconsecutive photon of a pair can be chosen by a so-called "correlation filter." In the process of WC, because SFG acts only as a spectral shift from telecom to visible wavelengths, it cannot exceed the original SNR with only shifting. Let us consider this effect from two perspectives: frequency and time domains. In the frequency domain, the degenerate-SPDC has a gain curve whose center of degenerate point shows the highest SNR per unit frequency. In this case, the wavelength converted range is positioned around the degenerate point, and it is possible to pick up the highly correlated two photons. In the time domain, in terms of noise, the main sources of noise before WC are the generation of new pairs of photons within the coherence time of another pair that has not yet completed its round trip in the cavity. Then, the decrease in $g^{(2)}_{s,i}(0)$ of telecom photons can be attributed to the probability of multipair generation in the cavity. After WC, although $g^{(2)}_{s,i}(0)$ depends on the SFG auxiliary laser power, resulting in WC efficiency and noise, the two photons that are separated in the time domain but are certainly correlated will be transferred to the visible range and their correlation will be enhanced within the spectral bandwidth of WC. This effect can also explain the result of very low SNR in one-photon WC: in the telecom range, there were many noise photons consisting of other pairs generated by next SPDC and, consequently, although the WC had high loss and additional noise, the correlation of the two photons reduced.

Although our WC crystal has a limited bandwidth of ~25 GHz that is restricted by the phase-matching condition, WC-TPC can achieve an adequate SNR despite attenuation by a dissonance between the bandwidths of WC and TPC. We demonstrated that the two-photon rate can be increased by controlling the SFG pump power owing to the correlation-filtering effect, which enables detection of photon pairs even if an uncorrelated photon exists within the coherence time of a pair. A trade-off exists between conversion efficiency and noise count depending on the auxiliary laser power. A lower noise count at an SPDC pump power < 1 mW is appropriate despite the decrease in conversion efficiency. To achieve an optimal conversion efficiency and noise count, a dedicated filter focusing on the TPC spectrum is required. We propose a filter containing $Pr^{3+}$:YSO with a high dopant concentration. Using rare-earth ions as a dynamical bandpass filter is not a new idea [50][51]; however, $Pr^{3+}$:YSO with a high dopant concentration of ~1% has not been studied well, and is expected to absorb a broad noise spectrum. Although we consider that QM absorption also plays a role in decreasing the noise from outside the two-photon spectra, it degrades the atomic frequency comb because ions may transfer from other hyperfine states.

After developing a special filter, WC can be customized as a polarization-insensitive type [39], which is difficult to achieve with the same efficiency as that of a single-polarization type. Polarization-insensitive WC is beneficial in the case of randomly rotating polarization when the conversion efficiency is more than half that of single-polarization WC. Irrespective of the type, the flying qubits should be in a time-bin state that is resistant to polarization changes along the transmission path [52]. Our single-polarization type WC can be directly applied to real quantum communication by using time-bin states because of the advantages of relative simplicity and higher efficiency: the polarization of time-bin photons will become unpolarized along with long-fiber transmission, and the condition will become the same as in this experiment, except for the quantum state.

In summary, we demonstrated VES with telecom TPC and wavelength conversion of two photons with a total fiber distribution length of 20 km. This technique renders a very narrow linewidth in the telecom range and high fidelity even with a long coherence time and multi-frequency modes, thereby creating prospects for quantum applications such as quantum networks combined with WC and the correlation-filtering effect. For further improving the TPC, a new method for separating two degenerate photons is desired to increase the distribution rate and investigate photon statistics. In the case of WC, a specialized filter is required to increase the SNR and establish a good relationship with QM. In addition, Bell state measurements for multimodes are preferable for frequency-multiplexing quantum communication applications. Our future plan is to convert polarization basis to time-bin basis and to add the loss of basis-conversion system to the WC-TPC result, which will enable us to estimate the whole quantum-communication system performance.

Methods

*Two-photon comb*

Two photons were generated by the SPDC process in two PPLN crystals (manufactured by Jinan Institute of Quantum Technology) having the dimensions of $0.5 \times 3 \times 10$ mm and $0.5 \times 0.5 \times 10$ mm, arranged orthogonally in a bow-tie cavity. The small dimensions allowed the crystals to precisely align with the laser path. The temperatures of the crystals and their holders were stabilized to be within ~1 mK. An SPDC pump laser of 757 nm was obtained by second harmonic generation (SHG) of a 1514-nm external-cavity diode laser (Sacher, TEC420-1530-1000), whose wavelength was stabilized using acetylene molecules. The pump laser was focused to a waist size of ~25 μm using lenses and a plano–concave mirror to achieve strong parametric interaction [53].

The optical cavity was stabilized using the Pound–Drever–Hall technique for this laser with an optical chopper with a duty cycle of 1/3. The two photons were separated using a 50:50 laser-line beam splitter, following which they entered a tomographic setup consisting of a zero-order 1514-nm quarter-wave plate, half-wave plate, and vertical-transmittance polarizer, similar to that in Ref. [43]. To generate a Bell state, two additional half-wave plates were placed in the path of one photon: one plate was used as a phase shifter by aligning the yaw angle, and the other was used as a bit flipper with a slow axis of 45°. The measurement time for 1 basis was 15 s, which was sufficient to converge the correlation function with a relatively strong pump power of 100 μW. The total testing time was ~10 min, including a rest time of 20 s. In almost all our experiments, the SSPDs were superconducting single-photon detectors with a detection efficiency of ~85%, which was the maximum value for our setup. However, a detection efficiency of ~60% was used in the experiments yielding the results shown in **Fig. 4 (b)** (blue dots) because our SiAPDs had an efficiency of 60% for visible wavelengths (SPCM-AQRH-14-FC). The maximal input power was 10 mW, resulting in a detected count rate approaching the limit of ~$10^7$. We utilized HydraHarp 400 as a TCSPC module, whose resolution is 32 ps for that shown in **Fig. 2 (a)**, and 16 ps and for the others. The 32 ps resolution was used because it could record longer interval times, which was required for measuring the coincidences with an adequate margin.

*Wavelength converter*

By removing the beam splitter and the tomographic setup, telecom photons were coupled to a polarization-maintaining fiber to be transported to the wavelength converter. An output collimator, consisting of a triplet lens, was placed immediately after the fiber for producing a high-quality single-mode Gaussian beam of diameter ~3 mm. The PPLN for the wavelength converter was of the waveguide type with an area of 9.9×11 μm and length of 48 mm (NTT electronics) with a type-0 phase matching condition (all three interacting lights are vertically polarized). The auxiliary laser for SFG has a wavelength of 1010 nm (TOPTICA, TA pro), where the emitted SHG light was stabilized using iodine molecules. Telecom photons and an auxiliary laser were coupled to this waveguide with efficiencies of ~60%. The external conversion efficiency was calculated by multiplying the coupling efficiency and the internal conversion efficiency. The external conversion efficiency was ~60%, whereas the internal counterpart was ~96% (details are presented in the Supplementary note 4). To obtain higher efficiency, we examined a lot of experimental conditions like changes in the combination of lenses, careful observation of waveguides by using camera, and optimization of noise filters. We found the best setup and realized the wavelength conversion of both two-photons. Two dichroic mirrors and one bandpass filter were installed as filters to remove WC noise, because using a spectrometer or numerous filters can cause optical loss, decreasing the two-photon coincidence count. In the process of obtaining WC-TPC, one-photon wavelength-converted coincidences were highly noisy that only the noise floor or an extremely low $g_{s,i}^{(2)}(0)$ was observed because of an excessively high photon rate at the telecom detector (~10 Mcts/s) due to pump-power-dependent uncorrelated photons.

Acknowledgements

We thank H. Goto, Q. Zhang, Y. Yamamoto, S. Utsunomiya, T. Kobayashi, M. Fraser, I. Iwakura, S. Tamura, K. Ikeda, and F.-L. Hong for their support. This work was supported by the Toray Science foundation, the Asahi Glass Foundation, the KDDI Foundation, the SECOM Foundation, Research Foundation for Opto-Science and Technology, JST PRESTO JPMJPR1769, JST START ST292008BN, and Kanagawa Institute of Industrial Science and Technology (KISTEC). TH also acknowledge members of Quantum Internet Task Force, which is a research consortium to realize the Quantum Internet, for comprehensive and interdisciplinary discussions of the Quantum Internet.

Author contributions

K.N. and T. H. conceived this project. K.N., D.Y., I.N., N.T., K.O., and T.H. designed the experiments. M.-Y.Z. and X.-P.X. fabricated the SPDC crystals. K.N., D.Y., and K.I. performed the experiments. K.N. and T.H. analyzed the data and drafted the manuscript. K.N., N.T., X.-P. X., and T.H. revised the text. All the authors contributed to discussions.

Data availability

The data that support the findings of this study are available from the corresponding author upon reasonable request.

Competing Interests

We declare that none of the authors have competing financial or non-financial interests as defined by Nature Research.

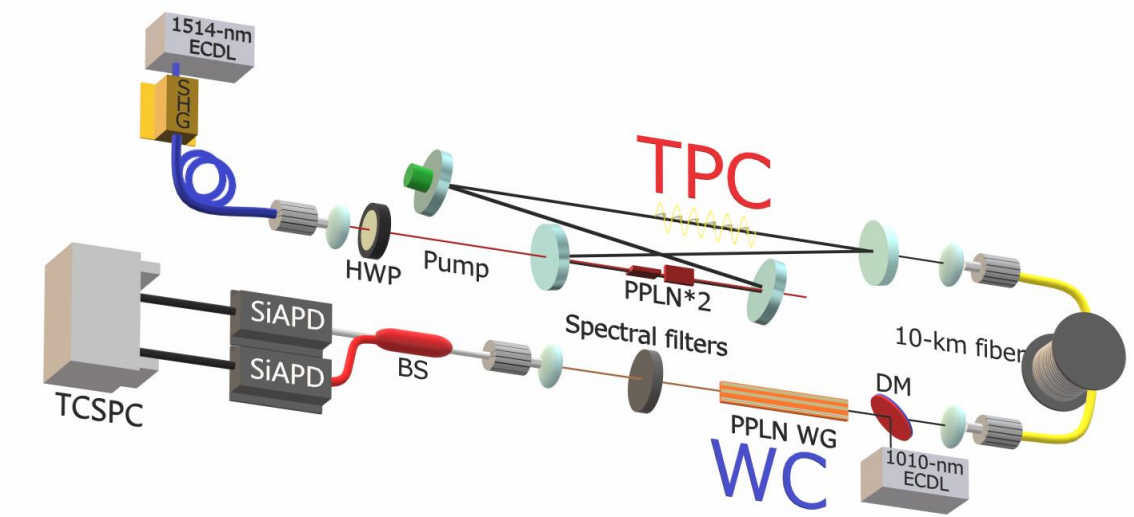

**Fig. 1** Experimental setup. The upper part shows two-photon comb (TPC), and the lower part shows a wavelength converter. SPDC pump light was produced by second harmonic generation (SHG) from a 1514-nm external cavity diode laser (ECDL). A half-wave plate (HWP) changes the polarization of the pump light to adjust the ratio of horizontal to vertical generation. SPDC occurs in two mutually orthogonal periodically poled lithium niobite (PPLN) crystals. After coupling to a single-mode optical fiber and transmitting through the 10-km fiber, two collimated photons are combined with a 1010-nm auxiliary laser using a dichroic mirror (DM) to convert the wavelength to 606 nm in the PPLN waveguide (WG). Spectral filters are composed of two dichroic mirrors and one bandpass filter. Detection is performed by two silicon avalanche photodiodes (SiAPDs) and a time-correlated single-photon counting module (TCSPC) with a fiber beam splitter (BS) to realize a Hanbury–Brown–Twiss-type interferometer. To perform two-photon state tomography, a long-pass filter and a 50:50 laser-line beam splitter are inserted before the 10-km fiber, and the two split photons pass through the set of a zero-order 1514-nm quarter-wave plate, a half-wave plate, a vertical-transmittance polarizer, and a fiber coupler to be detected by superconducting single photon detector (SSPD) of superconducting single photon detector.

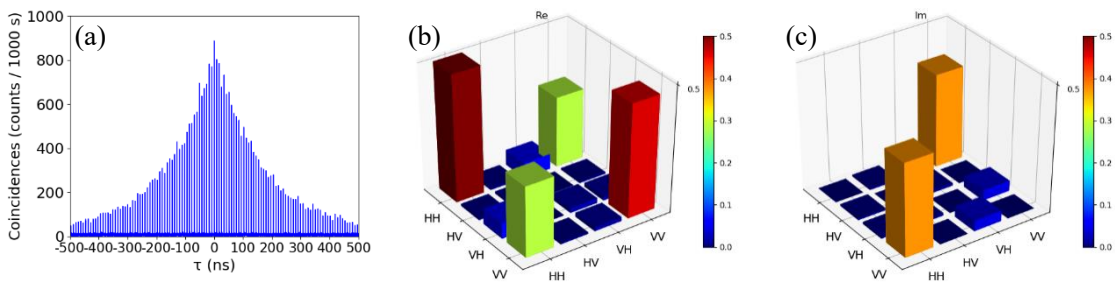

**Fig. 2** Experimental results of the telecom two-photon comb (TPC). (a) Glauber's two-photon correlation function. The time interval is a cavity round-trip time of 8.6 ns, and the free spectral range (FSR) can be reciprocally estimated as ~120 MHz. The exponential-envelope function had a cavity

linewidth of 0.95 MHz, which showed a coherence time of up to 1 μs. Reconstructed density matrices showing the absolute value of (b) real part and (c) imaginary part. H: horizontal, V: vertical.

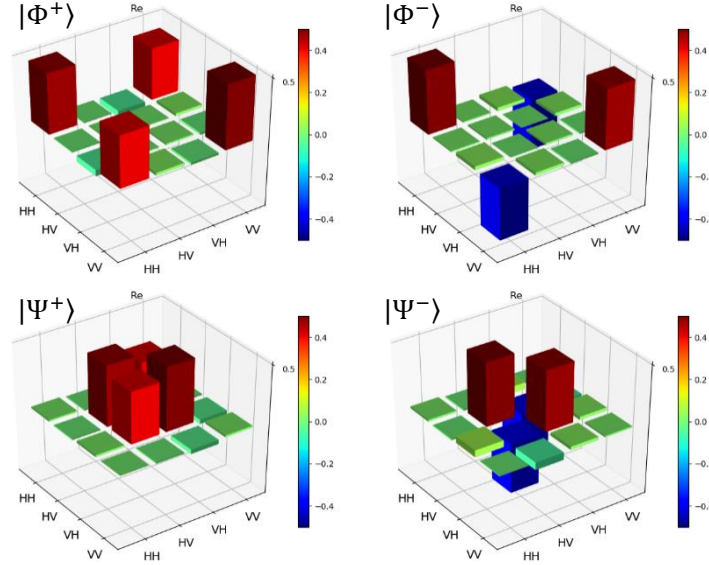

**Fig. 3** Reconstructed Bell states: $|\Phi^+\rangle, |\Phi^-\rangle, |\Psi^+\rangle, |\Psi^-\rangle$. Only the real parts are shown because all the imaginary parts were nearly zero. The fidelity to $|\Phi^+\rangle = |HH\rangle + |VV\rangle$, $|\Phi^-\rangle = |HH\rangle - |VV\rangle$, $|\Psi^+\rangle = |HV\rangle + |VH\rangle$, and $|\Psi^-\rangle = |HV\rangle - |VH\rangle$ was 90.0%, 90.2%, 89.4%, and 88.1%, respectively (we omitted the coefficients $1/\sqrt{2}$). The raw counts are presented in Supplementary note 1.

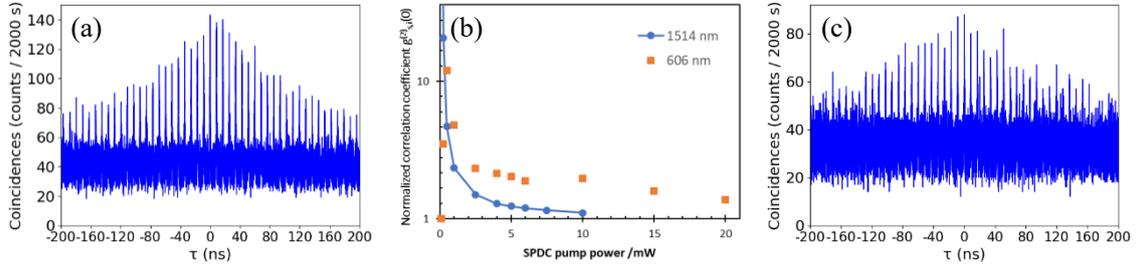

**Fig. 4** Experimental result of wavelength-converted two-photon comb (WC-TPC). (a) WC-TPC without 10-km optical fiber. Measurement conditions: SPDC pump power of 10 mW; SFG auxiliary laser power of 50 mW; silicon avalanche photodiode (SiAPD) efficiency of ~60%. (b) Normalized correlation coefficient $g_{s,i}^{(2)}(0)$ versus SPDC pump power. The blue circles connected by a line represent 1514-nm two-photon data (without wavelength conversion), and the orange squares represent the WC-TPC data. The error bars, which mean 1 standard deviation of $g_{s,i}^{(2)}(0)$, are smaller than the points (or some can be seen in lower power range; see Supplementary Figure 2. (b)) and are therefore omitted. Although superconducting single photon detector (SSPD) and SiAPD have almost the same efficiency of 60% owing to the adjustment of applied current, the timing jitters were very different (~40 ps for SSPD, and ~300 ps for SiAPD). (c) WC-TPC with 10-km fiber. Measurement conditions are same as Fig. 4 (a).

Supplementary Information

Two-photon comb with wavelength conversion and 20-km distribution for quantum communication

Supplementary note 1: Quantum state tomography

As described in the manuscript, we measured two-photon coincidence counts in 16 polarization basis. This follows Ref. [43] and demands only one waveplate rotation between each measurement, shortening the total measurement time. **Supplementary Figure 1** shows coincidence counts for the Bell states $|\Psi^+\rangle$ and $|\Psi^-\rangle$, which are realized by installing 45° slow-axis half-wave plate without yaw angle and 0° slow-axis half-wave plate with a yaw angle. The apparent difference appears in the counts for n10 ($|DD\rangle$) and n16 ($|RL\rangle$): By changing the yaw angle of only one half-wave plate, these counts visibly increase and decrease. Because of this advantageous nature, we could achieve high fidelity and arbitral relative-phase adjustment. The counts for $|\Psi^-\rangle$ is a slightly low because of the misalignment of the optical path due to larger yaw angle.

Tomographic reconstruction and maximum likelihood method are achieved based on Ref. [2], which is an available program in Git Hub; thus, there is no room for our conveniences or mistakes in tomography experiment. Furthermore, to depict the density matrix, QuTiP library in python is used [3]. In this whole processing, fidelity is defined as $\mathcal{F} = \langle\phi_{ideal}|\rho_{reconstruct}|\phi_{ideal}\rangle$, which is a common definition (however, this is not square-root fidelity [4]). For further higher fidelity, improvement of the basin-like structure, which appears small in n16 count (left side of **Supplementary Figure 1**), is required. This is caused by polarization rotating along cavity-round-trip phase shift by passing through crystals or reflection on dielectric mirror with a few angles; it can be reduced by finely adjusting the crystal position and decreasing the reflection angle.

Our main effort aiming at high fidelity is to reduce decoherence effect as much as possible. We construct a bow-tie cavity with a small reflection angle (approximately 2-3 degrees) to decrease senkrecht and parallel polarization dependence. This angle is limited by finite width of PPLN crystal; the minimum magnitude is proportional to arctangent of the half of the short arm of bow-tie configuration and the width of PPLN. Long cavity can make reflection angle smaller than ordinal-size cavity, resulting in decoherence. Also, a PPLN crystal has birefringence, and we adjust the temperature of two PPLNs with a few milli-kelvin stabilities to reduce total birefringence. Furthermore, the mirrors after the cavity for coupling to the PMF are all removed, and two photons straightly couple into the PMF. These are only a few examples to realize high fidelity: dephasing mechanism is very complex in reality.

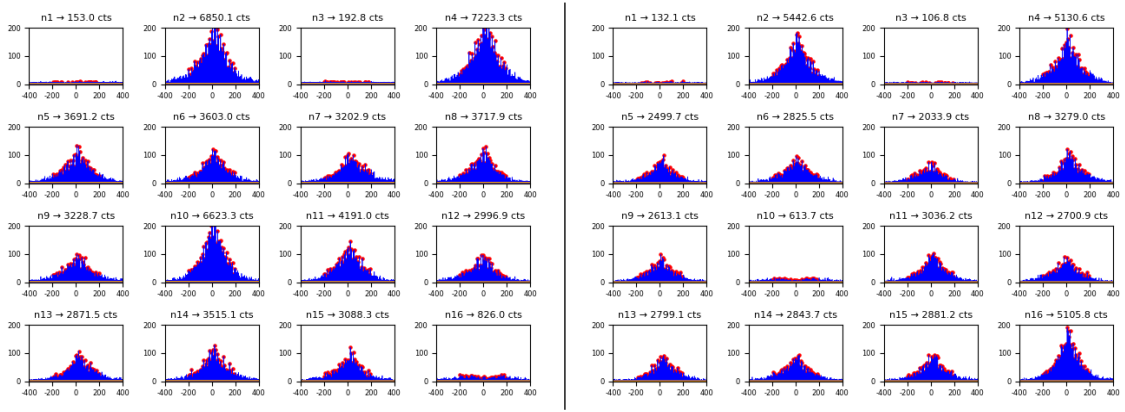

**Supplementary Figure 1.** Two examples of coincidence count in 16 measurements, on the $|HH\rangle, |HV\rangle, |VV\rangle, |VH\rangle, |RH\rangle, |RV\rangle, |DV\rangle, |DH\rangle, |DR\rangle, |DD\rangle, |RD\rangle, |HD\rangle, |VD\rangle, |VL\rangle, |HL\rangle, |RL\rangle$ basis. Left (right) is the case of generating Bell state of $|\Psi^+\rangle$ ($|\Psi^-\rangle$). Blue represents raw counts and red represents counts within a range of [-200, 200] ns. SPDC pump light is nearly $|D\rangle$ polarized with a power of 100 μW. Each acquisition time is 15 s, and the total time becomes about 10 min due to waveplate rotation. The only difference between these two experiments is the yaw angle of the half-wave plates.

Supplementary note 2: Dependence of two-photon coincidences on SPDC pump power

We demonstrated that $g^{(2)}_{s,i}(0)$ of WC-TPC can overcome that of original TPC (Fig. 4 (b)). In this supplementary note, we focus on original TPC or telecom two-photon correlation before WC. **Supplementary Figure 2** (a) shows two-photon timing correlation with the conditions of SPDC-pump power of 10 mW, polarization of only |H⟩, SSPD detection efficiency of ~ 60%, and bin size corresponding to the resolution of 16 ps. Noise slope can be observed at the right side in this figure, which continues to plummet which very rapidly reach the zero count. This is natural for the case of high-count range at nearly 10 MHz because coincident detection occurs in the vicinity of zero. Although this noise curve must have continued toward the left (minus of time axis), it appears flat for some range. This results from the dead time ~ 80 ns of our time-correlated single photon counting module HydraHarp 400. Without dead time, the whole noise function would become a slope and $g^{(2)}_{s,i}(0)$ would be different from this data.

**Supplementary Figure 2** (b) is a log scaled version of Fig. 4 (b), which is obtained by only changing SPDC pump power (all the other parameters are the same as above). TPC without WC can be observed more clearly with higher $g^{(2)}_{s,i}(0)$ in lower power range. This is because the photon-pair number inside a cavity becomes lower and the noise caused in the crystal approaches zero. This function must converge to 1 at very low pump power because no pump would cause only dark counts; however, we have never observed this point because our SSPD has excellent-low noise count (in the order of 10 cps). In the low power range, the gradient in the logarithmic graph shows almost -1 resulting in inversely proportional change to the pump power.

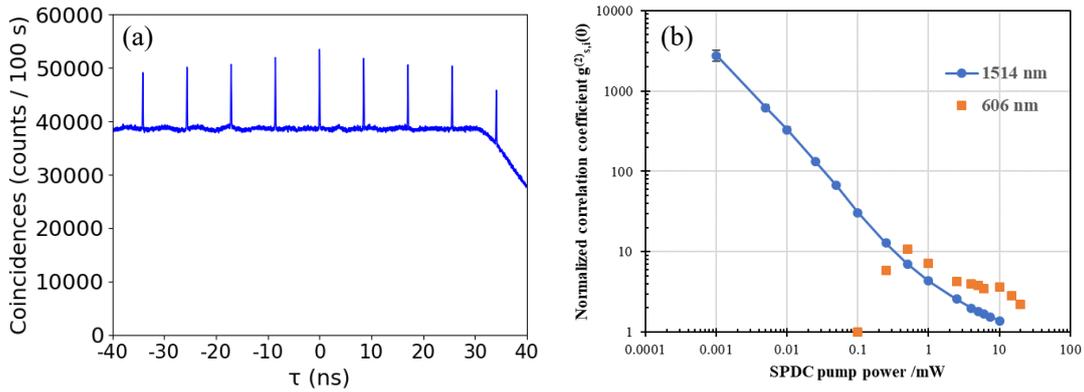

**Supplementary Figure 2.** Two-photon coincidences focused on SPDC pump power. (a) Coincidence counts of 1514 nm two photons. SPDC pump power is 10 mW, and the detection efficiency of superconducting single photon detector (SSPD) is adjusted as ~ 60% to not reach the maximum rate per channel. (b) Normalized correlation coefficient $g^{(2)}_{s,i}(0)$ versus SPDC pump power (same as Fig. 4 (b) in main article). The blue circles connected by a line represents 1514 nm two photon data (without WC), and the orange squares represent WC-TPC. Although SSPD and silicon avalanche photodiode

(SiAPD) have almost the same efficiency of 60% because of regulation of the applied current, the timing jitters differ greatly: ~40 ps for SSPD and ~300 ps for SiAPD.

Supplementary note 3: Estimation of bandwidth of TPC spectrum

In general, spectrometer or optical grating is widely used to analyze a spectrum. However, these "bandpass method" has a little difficulty of high optical loss. Further, when measuring both two photons, two frequency-correlated setups are required because two photons satisfy energy conservation. Then, we suggest another new method to estimate two-photon spectrum with lower loss and relatively easy setup.

We utilize 10 km single-mode fiber and observe dispersion of TPC. In general, optical fiber has birefringence and the optical path length could change depending on the fiber condition. When only the influence of dispersion is required, a correlation of two photons dispersed in the same fiber is suitable under the approximation that the fiber remains almost unchanged within the coherence time. **Supplementary Figure 3** shows dispersed two-photon correlation after 10 km fiber with an SPDC pump power of 10 µW. These results are obtained due to a long time interval caused by our long cavity. To estimate the timing width, we used the following equations:

$$G_{s,i}^{(2)}(\tau) = c \times e^{-2\pi f_{FWHM}|\tau|} \times \text{Comb}(\tau) + \text{noise},$$

$$\text{Comb}(\tau) = \sum_{n=-(N-1)}^{N-1} \exp\left[-4(\ln 2)\left(\frac{\tau - n t_{FSR}}{T_{FWHM}}\right)^2\right],$$

where $c$ is a coefficient corresponding to peak counts; $f_{FWHM}$ represents full width at half maximum (FWHM) of one frequency peak, i.e., linewidth; $N$ is related to the number of timing combs; $t_{FSR}$ denotes the time interval expressed as a reciprocal of free spectral range (FSR); $T_{FWHM}$ is the FWHM of one timing peak which is approximated by a Gaussian-shape peak. Although a Voigt function should be used for the strict estimation of the spectrum, we used a Gaussian function to reduce complexity. The red line corresponds to this function; although it appears smaller than raw data, the insets show that it has some variance around the peak counts. From this figure, despite passing through the 10 km fiber, frequency linewidth $f_{FWHM}$ is almost the same as that of ~1.35 MHz that corresponds to 95% reflectivity of cavity output coupler. $T_{FWHM}$ is $2.0 \pm 0.2$ ns ns from (a) and $4.0 \pm 0.3$ ns from (b) among 10 peaks. The timing jitter of our measurement system is ~0.1 ns, and has ignorable influence on these values. The dispersion value around 1514 nm can be calculated using the dispersion equation as ~+15 ps·nm$^{-1}$·km$^{-1}$; then, we estimate the TPC frequency bandwidth of ~13 nm.

Such a very long timing can be compensated using a special fiber having an opposite sign of dispersion value. If this compensation can be considered as one of the re-modulations of wave packet, it can recover the quantum coherence as well [6]. Moreover, using band-pass effects such as WC, we can eliminate this broadening (Fig. 4 (c) in main text).

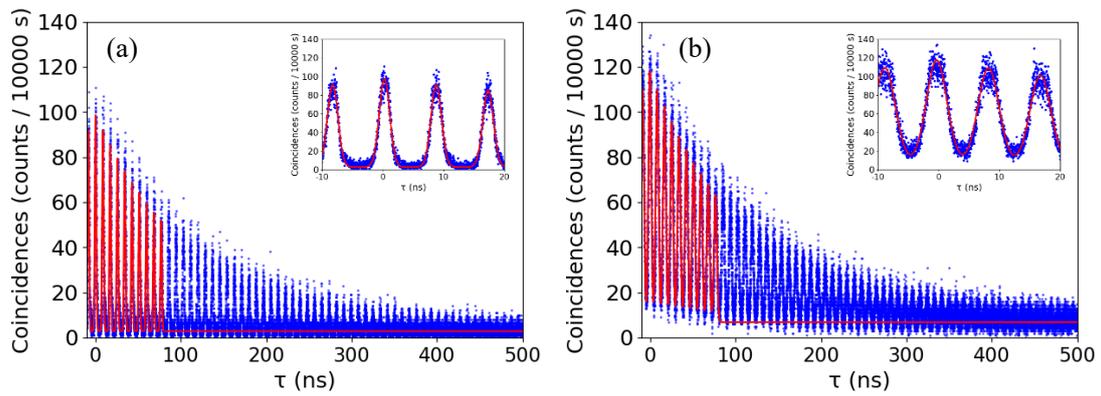

**Supplementary Figure 3.** Two-photon coincidence counts with dispersion after passing through the 10 km optical fiber: (a) only one photon and (b) two photons. Blue represents raw data point and red represents the fitted function (only 10 pieces of gaussians are pictured). Insets are the enlargements of the regions around the center $\tau = 0$.

Supplementary note 4: The efficiency and noise of wavelength converter

The main characteristics of a wavelength converter are the quantum efficiency of conversion and the counts of noise. Quantum efficiency is calculated as follows [7]:

$$\eta = \frac{N_{converted}}{N_{telecom}} = \frac{P_{convered}/h\nu_{converted}}{P_{telecom}/h\nu_{telecom}} = \sin^2\left(L\sqrt{\eta_{nor}P_{pump}}\right),$$

where $N, P, \nu, L, \eta_{nor}$, and $P_{pump}$ represent the number of photons, power of light, frequency of light, length of wavelength-conversion crystal, normalized power efficiency in the low-gain limit, and pump power, respectively. We used a 48 mm PPLN waveguide with a cross section of 9.9 μm × 11.5 μm to realize higher efficiency than bulk.

**Supplementary Figure 4** (a) shows a quantum efficiency $\eta$. We used telecom laser to directly measure the external efficiency, which is the conversion efficiency, including waveguide-coupling efficiency and transmission loss, but removed the influence of the auxiliary laser. Further, we obtained the telecom-laser coupling rate by turning the WC pump laser off, to estimate the internal efficiency. The maximum external efficiency we could measure was 56.0%; however, it appears to increase slightly at higher pump power range. The coupling rates at 1514 nm and 1010 nm are 59.4% and 60.5%, respectively; then, we can estimate the internal efficiency. Our crystal is designed to have an efficiency of nearly 1, and the measured maximum value was 94.3%.

Noise count is obtained as shown in **Supplementary Figure 4** (b). In this measurement, only WC pump laser is used, and the filters consist of two dichroic mirrors and one bandpass filter. The dotted curve represents quadratic-function fitting and the measured data set are almost positioned on fitting. This indicates that noise is mainly caused by second-order nonlinear optical process in this setup. Then, we analyze the component noise spectrum using a spectrometer. The result indicates one peak, which is centered on 606 nm and a FWHM of ~ 25 GHz (not shown). We then try to adjust the number of filters or optical depth by adopting other bandpass filters or filtering crystals.

In **Supplementary Figure 5**, we defined the wavelength-conversion function $\eta^*$ as the ratio of external efficiency and noise rate at kilo counts per second. In this process, the approximated function $\eta$ and the quadratic noise function are used. This function has no meaning in absolute value because of strange unit (% divided by kilo counts), but indicates the pump-dependent signal-to-noise ratio. This calculation is similar to reference [8]. The maximum point is positioned at 7.97 mW pump power in our condition of dark counts of ~0.32 kHz; on using lower dark counts, this point shifts toward the left. Although this pump power was the best to obtain high SNR, we chose a higher power of 50 mW to obtain shorter experimental time for accurate comparison of $g_{s,i}^{(2)}(0)$ (**Supplementary Figure 2** (b)). As one of the indices, noise equivalent power (NEP) is often adopted [9], and the optimum pump power of that shows 46.8 mW. The actual comparison of the two-photon coincidences is shown in the next supplementary note.

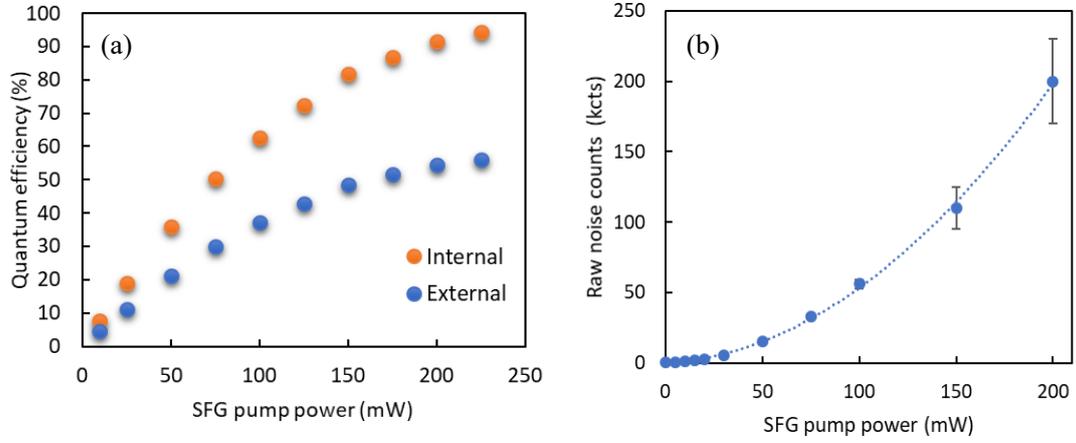

**Supplementary Figure 4.** Specifications of wavelength converter. (a) Conversion efficiency, defined as the ratio of the number of photons before and after WC. (b) Raw noise counts obtained by inputting only 1010 nm auxiliary laser and installing two dichroic mirrors and one bandpass filter.

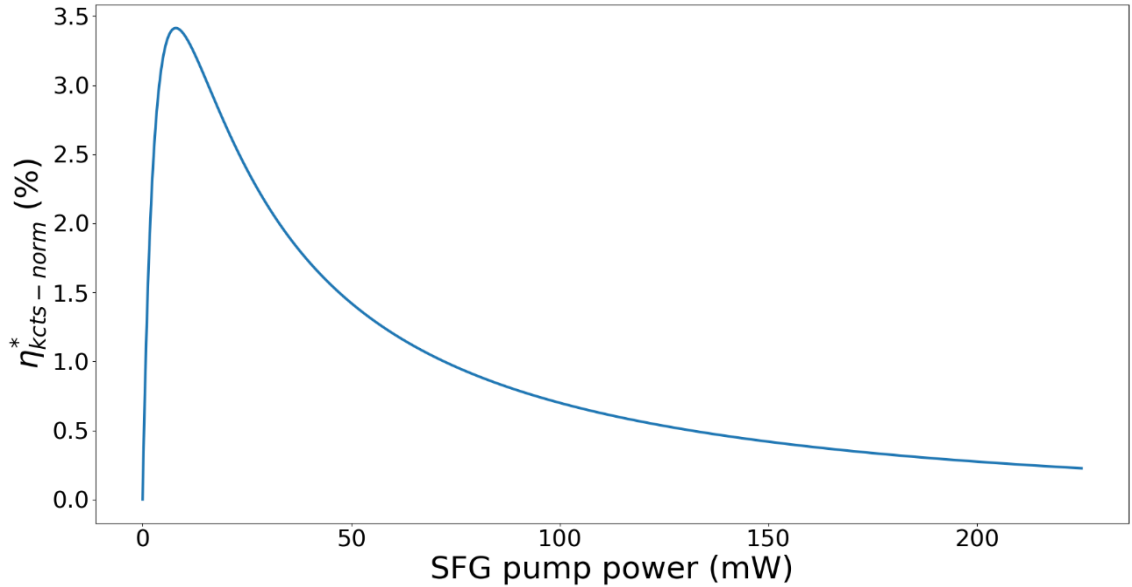

**Supplementary Figure 5.** Noise-normalized conversion efficiency $\eta^*$, which we defined as the external efficiency curve divided by the noise count curve of kilo counts per second. The peak can be realized at the SFG pump power of 7.97 mW. This value is strongly influenced by the dark count rate.

Supplementary note 5: Dependence of WC-TPC on auxiliary laser power

This note is related to wavelength conversion efficiency and noise count. In the previous section, we demonstrated that a trade-off exists between them, and we had to search for a balanced pump power of WC. **Supplementary Figure 6** shows pump-power dependent $G_{s,i}^{(2)}(0)$ of WC-TPC with SPDC pump power of 5 mW. The values in the figure correspond to external conversion efficiency and noise level (raw count rate, i.e., SPD detection rate after passing filters and coupled to single-mode fiber without beam splitter). Because the heights of peak are arranged to almost same among the figures, the noise floors correspond to normalized correlation functions $g_{s,i}^{(2)}(0)$. These three functions indicate that lower pump power corresponds to better SNR. In our setup, telecom two photons, whose counts are stabilized within 10% of the accumulating time, are transmitted through a polarization-maintaining fiber; therefore, converted photons must be of almost the same number. Further, the main reason is the increase in noise count with the increasing pump power.

We compare **Supplementary Figure 7** (a) and (b) to perform WC-TPC with lower pump power: (a) is the case of very weak SPDC pump of 100 μW and WC pump of 10 mW, whereas (b) has the same SPDC pump but a WC pump of 50 mW. There is a clear difference in $g_{s,i}^{(2)}(0)$, but the coincidence rate of (a) is very low at ~0.1 cts/sec (3000 cts/30000 sec).

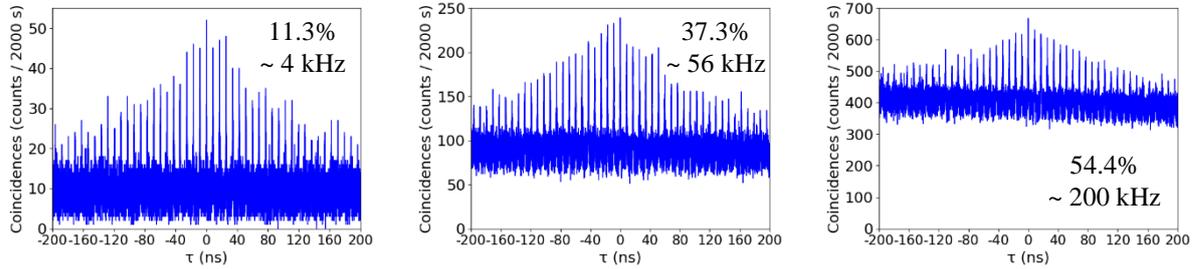

**Supplementary Figure 6.** WC-TPC at various WC pump powers of 25 mW, 100 mW, and 200 mW. These all are used 5-mW SPDC pump power. The values shown inside the plots are external quantum efficiency (top) (%) and noise count rate (bottom) (kHz).

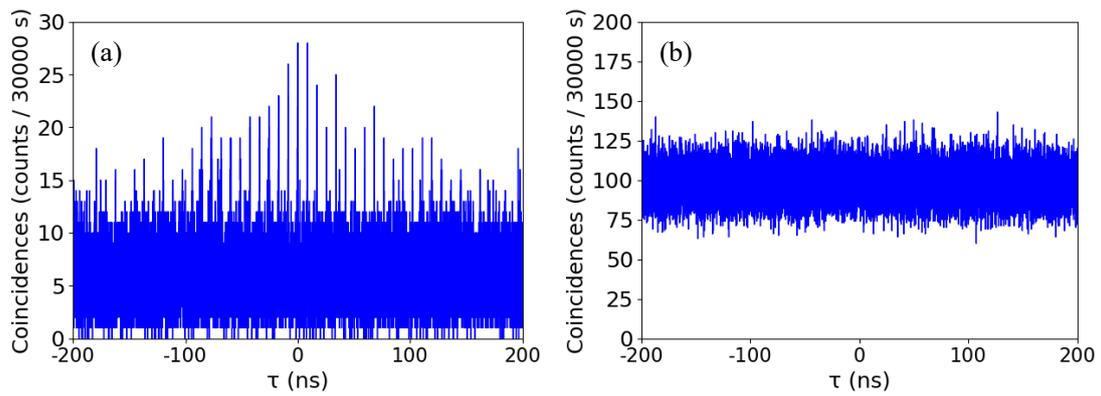

**Supplementary Figure 7.** (a) WC-TPC with relatively weak lasers. SPDC pump is 100 μW and WC pump is 10 mW. At higher WC pump power, this structure cannot be realized because of noise. (b) Same SPDC pump as in (a) but WC pump of 50 mW.

Supplementary note 6: Comparison with previous studies on narrow-linewidth photon sources

It must be beneficial to compare with previous researches. We would like to emphasize that there is no demonstration of telecom narrow-linewidth entanglement source. **Supplementary Table 1** lists some excellent photon sources with narrow linewidth. Although it requires spectral brightness to evaluate the photon source, a more important index is the operational photon-generation rate, which can be enhanced by increasing the pump power up to the parametric threshold. Ref. [15] performed multi-mode polarization-entanglement state tomography; however, it indicates approximately 10% worse fidelity than the single-mode case. This shows a possibility that our photon source can realize higher fidelity with a fewer number of modes. Furthermore, narrower linewidth corresponds to longer coherence time; thus, realizing high fidelity becomes difficult because a wider timing window is required, resulting in decreased fidelity in accordance with the imperfection of experiment like instability or accidental coincidences [6]. The TPC setup is suitable to overcome degradation of quality caused by experimental imperfections and long time passing, for instance, maintaining a sufficiently narrow linewidth and realizing high fidelity.

| 1st Author (Year) | Characteristics | λ (nm) | Δν (MHz) | Fidelity (%) | Mode |
|---|---|---|---|---|---|
| [10] J. Fekete (2013) | One wavelength aims at direct absorption of $Pr^{3+}$:YSO | 1436 606 | 1.7, 2.9 | - | 1 |
| [11] Z. -Y. Zhou (2014) | Telecom | 1560 | 8 | - | Multi |
| [12] K. Liao (2014) | Four wave mixing using $^{85}$Rb with magneto-optical trap | 780 795 | 0.8 | 95.2 | 1 |
| [13] L. Tian (2016) | Single Mode (Degenerate SPDC) | 795 | 15 | 95.2±0.8 | 1 |
| [14] M. Rambach (2016) | Flip-trick technique to realize very narrow linewidth | 795 | 0.67* | - | Multi |
| [15] J. Wang (2018) | Alignment using optical wedges | 935 880 | 9, 9.5 | 89.6 (~80) | 1 (Multi) |
| **This work** | **Telecom narrow linewidth and entanglement with frequency multimode** | **1514** | **0.95** | **96.1** | **Multi** |

**Supplementary Table 1.** List of photon sources. Although numerous photon sources have been developed to date [16], only a few examples were given from the standpoint of outstanding value. * this value indicates the cavity damping rate, although they assert that the actual linewidth of photon (or 1 mode) must be 0.64 times smaller than the damping rate.

Supplementary note 7: Atomic frequency comb quantum memory

This note supplies the outline of atomic frequency comb (AFC) quantum memory scheme using $Pr^{3+}$:YSO [17]. $Pr^{3+}$ is the dopant of YSO crystal and the concentration corresponds to replacement percentage of Y atom with $Pr^{3+}$ ion. Then individual $Pr^{3+}$ ion possesses specific absorption line around ~ 494.7 THz or 606 nm because of surrounding environment or crystal field. As a whole ensemble, this crystal looks having wide-broadened absorption line, as named, inhomogeneous broadening (see **Supplementary Figure 8** right). Inhomogeneous broadening has a width of the order of from GHz to hundred GHz depending on dopant concentration. AFC structure is tailored into inhomogeneous broadening within a range of < 4 MHz, the width of < 100 kHz, and the comb separation of ~ 100 kHz (this separation is directly related to rephasing time or photon echo time). For instance, the width of (36±3) kHz and comb separation of approximately 137 kHz are used [18]. This comb structure can be established thanks to hyperfine structure (see **Supplementary Figure 8** left) and the long lifetime of that. Because AFC structure is some-order narrower than inhomogeneous width, we can tailor many AFC structure [19][20] with separation of ~ 120 MHz, which corresponds to the FSR of present two-photon spectrum (**Supplementary Figure 9**). Our wavelength converter has bandwidth of 25 GHz and it just fits to inhomogeneous width of 1-100 GHz. On the other hand, our (telecom) TPC has ~1-THz overall bandwidth and if it becomes necessary to use more than 25 GHz for more frequency multiplexing, wider bandwidth wavelength converter is required.

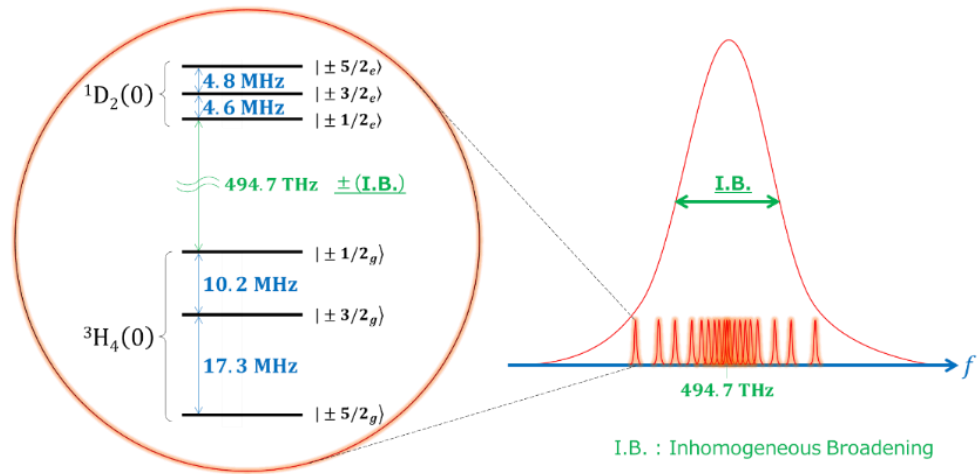

**Supplementary Figure 8.** Energy diagram (left) and inhomogeneous broadening (right) of $Pr^{3+}$:YSO.

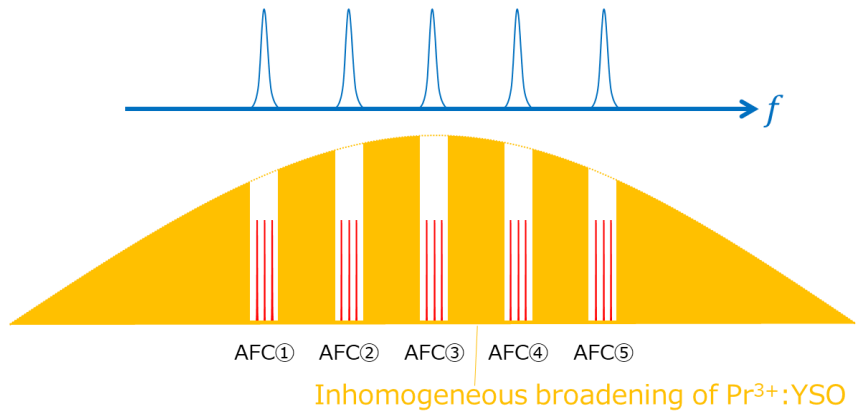

**Supplementary Figure 9.** AFC structure for frequency multimode.